# Surface Reduction Boosts Free Electron Concentration in MXene for Enhanced Photothermal Performance

*Haoming Ding[1,2#], Xiao Tong[1,2#], Yong Zhang[1,2,3]\**.

[1]*Department of Biomedical Engineering, College of Biomedicine, The City University of Hong Kong, Hong Kong SAR, China*

[2]*Cardiovascular and Cerebrovascular Health Research Centre (COCHE), Hong Kong SAR, China*

[3]*Department of Materials Science and Engineering, College of Engineering, The City University of Hong Kong, Hong Kong SAR, China*

[#]These authors contributed equally.

[*]To whom correspondence may be addressed. E-mail: yozhang@cityu.edu.hk

**Abstract:** The photothermal properties of MXenes originate from their high free electron concentration, which drives localized surface plasmon resonance (LSPR). However, their intrinsic electron concentration is limited by suboptimal *d*-orbital occupancy, while electronegative surface terminations actively deplete free electrons through orbital-selective withdrawal. Herein, we report a sodium (Na)-mediated surface reduction strategy in molten salts to transform $Ti_3C_2Cl_x$ into electronically tunable $Ti_3C_2$. Specifically, Na atoms remove -Cl terminations to eliminate electron withdrawal and simultaneously inject electrons into the MXene lattice via a reduction reaction. This dual effect saturates Ti 3*d*-orbital vacancies while reducing surface coordination sites, achieving an increase in carrier concentration to 4.92-fold, an increase in mobility to 2.63-fold, and an enhancement in conductivity to 12.96-fold. Consequently, the optimized reduced MXene achieves a record photothermal conversion efficiency of 92.36% under 808 nm laser irradiation. As a proof-of-concept, a photothermal antibacterial woundplast with ultralow MXene content demonstrates a 91.39% bacterial kill rate. This work not only shows an effective way to tune the photothermal properties of MXenes but also inspires applications that require high electron concentration, such as energy storage, sensors, and electromagnetic interference shielding.

**Introduction**

MXenes represent an emerging class of two-dimensional (2D) transition metal carbides and nitrides, typically synthesized by selectively etching the A layer (*e.g.*, etching Al from $Ti_3AlC_2$) using different etchants (*e.g.*, hydrofluoric acid (HF), LiF+HCl, and Lewis acidic molten salts).[1-5] This process yields materials with the general formula $M_{n+1}X_nT_x$, where M is an early transition metal (*e.g.*, Ti, V, Nb, and Mo), X is carbon/nitrogen, and $T_x$ denotes surface terminations (-O, -OH, -F, and -Cl) inevitably introduced during synthesis or subsequent processing.[2,5] This unique architecture endows MXenes with exceptional metallic conductivity, mechanical flexibility, hydrophilicity, and tunable electronic properties, driving application exploration across energy storage, catalysis, and biomedicine.[5-8] For biomedical applications, MXenes have attracted considerable attention due to their photothermal conversion capabilities, especially in the near-infrared (NIR) biological window.[9-12] This capability is primarily driven by localized surface plasmon resonance (LSPR), a phenomenon arising from the oscillation of free electrons in the conductive MXene layers when excited by NIR photons.[13,14] The *d*-orbitals of transition metals endow MXenes with a high density of states (DOS), which is essential for robust LSPR, thereby establishing the free electron concentration as a pivotal factor governing photothermal conversion efficiency (PTCE).[15,16] Currently, the PTCE of MXenes is still low, which limits their application in the biomedical field.[12] Many efforts have been made to improve their photothermal performance through structural and electronic engineering strategies.[13,17-19]

Strategies to improve the photothermal performance of MXenes include decorating the MXene surface with plasmonic noble metal nanoparticles (*e.g.*, Au, Ag, and Pt) to enhance NIR absorption.[17] However, this approach introduces complexity in synthesis and raises concerns about nanoparticle stability and biocompatibility. Other approaches include intrinsic electron modulation, such as nitrogen doping in $Ti_3C_2$ to increase DOS *via* additional electron donation, or designing high-entropy MXenes to induce *d*-orbital hybridization and delocalize electrons.[18,19] A more direct approach is surface chemistry tuning, as electronegative terminations extract electrons from the $M_{n+1}X_n$ lattice, thereby suppressing free electron concentration.[20-22] It is reasonable to assume that removing these surface terminations to create bare MXene could theoretically restore intrinsic DOS and PTCE.[22,23] However, complete, damage-free removal of surface terminations remains challenging. Conventional high-temperature annealing can only remove weakly bound surface terminations (*e.g.*, -F and -OH), while chemical reduction with sodium hydride (NaH), gallium (Ga), or hydrogen ($H_2$) usually results in incomplete termination removal or 2D lattice destruction.[20,24-26] On the other side, since the *d* orbitals of the M atoms in MXenes are not filled, external electron injection can significantly increase the electron concentration, which is a significant advantage over other two-dimensional materials such as graphene, boron nitride (BN), and transition metal chalcogenides (TMDs). However, to the best of our knowledge, this electron injection process has not yet been demonstrated.

Here, we report a sodium (Na)-mediated molten salt reduction strategy that achieves dual electronic optimization: (1) almost complete removal of electronegative surface terminations, eliminating their electron-withdrawing effect; (2) direct electron injection into the $Ti_3C_2$ lattice *via* Na-induced Ti reduction. This synergistic process achieves an increase in carrier concentration to 4.92-fold, an increase in mobility to 2.63-fold, and an enhancement in conductivity to 12.96-fold. As a result, the optimized reduced $Ti_3C_2$ achieves an ultrahigh PTCE of 92.36% under 808 nm laser irradiation, which is the highest efficiency reported for MXenes at this wavelength. As a proof-of-concept, a light-triggered woundplast with ultralow MXene content can effectively kill 91.39% of *Staphylococcus aureus*, highlighting the translational potential of this approach.

**Results and Discussion**

**Synthesis of Reduced $Ti_3C_2$ MXene**

Figure 1a illustrates the synthesis of reduced MXene via the reaction of $Ti_3C_2Cl_x$ with metallic Na. The pristine $Ti_3C_2Cl_x$ was prepared by etching $Ti_3AlC_2$ MAX phase using a Lewis acidic molten salt ($CdCl_2$), as previously reported.[3,4] This precursor was then reacted with Na under an inert atmosphere, where molten salt enhanced reaction kinetics and thermodynamics. In this process, Na donates electrons to electronegative -Cl terminations ($\chi \approx 3.16$ vs. Ti $\approx 1.54$), converting Cl into $Cl^-$ that diffuses into the molten salt (Equation 1). By adjusting Na content and temperature, electron injection was controlled (Fig. 1b, Table S1). With $x \approx 2$ in $Ti_3C_2Cl_x$, we first tested a 1:2 molar ratio at 500°C to remove -Cl terminations. Increasing Na up to 1:8, studied further electron injection (Equation 2). At 500°C, the structure remained intact; raising the temperature to 550°C with the same ratio further improved electron transfer. Products were labeled as reduced $Ti_3C_2$ (1:2, 500°C), (1:8, 500 °C), and (1:8, 550°C), respectively. However, structural degradation occurred at 600°C or a 1:12 ratio (Fig. S1), indicating the accommodation limit of injected electrons. As a result, optimal conditions for maximal electron concentration while preserving the layered structure are 550°C and a 1:8 ratio. This stepwise design not only enables precise control over the reduction degree but also provides insight into the thermodynamic and kinetic factors governing the electron injection mechanism in MXenes.

$$Ti_3C_2Cl_x + xNa = Ti_3C_2 + xNaCl \tag{1}$$

$$Ti_3C_2 + Na = [Ti_3C_2]^- + Na^+ \tag{2}$$

X-ray diffraction (XRD) patterns of the pristine $Ti_3C_2Cl_x$ and reduced $Ti_3C_2$ synthesized under different conditions are shown in Fig. 1c. XRD, scanning electron microscopy (SEM), and energy dispersive spectroscopy (EDS) results of $Ti_3C_2Cl_x$ annealed in eutectic LiCl-KCl molten salt at 500°C without Na metal show no significant structural and compositional change, eliminating the effect of temperature and molten salt (Fig. 1c and Figs. S2, 3). In contrast, reduced $Ti_3C_2$ samples synthesized under different conditions exhibit notable variations in the (002) peak, corresponding to interlayer spacing values of 1.209 nm ($Ti_3C_2$, (500°C, 1:2)), 1.159 nm ($Ti_3C_2$, (500°C, 1:8)), and 1.252 nm ($Ti_3C_2$, 550°C, 1:8)). These differences are attributed to the evolution of interlayer interactions during reduction. Upon removal of -Cl terminations, exposed Ti atoms form dangling bonds, generating strong interlayer Coulombic repulsion and increasing the spacing.[27] As electrons are injected, they occupy Ti $3d$ orbitals, screening the positive charges and enhancing in-plane metallic bonding, which reduces the interlayer distance.[28,29] However, under excessive electron injection, high-density negative charges accumulate between layers, leading to dominant interlayer Coulombic and Pauli repulsions, resulting in a sudden expansion of the interlayer spacing.[30] This evolution of interlayer spacing reflects an electronic-structure-driven transition in interlayer interactions, offering insight into the tunable structural behavior of MXenes.

Figure 1d, e shows SEM images of pristine $Ti_3C_2Cl_x$ and annealed $Ti_3C_2Cl_x$, respectively. Unlike HF-etched MXene, which typically exhibits a distinct accordion-like layered morphology, both particles show compact layered structures.[1,31] This is due to the strong interlayer binding caused by -Cl terminations.[3] In contrast, the particle of reduced $Ti_3C_2$ (1:8, 550°C) displays a more pronounced accordion-like layered structure, which aligns with the increased interlayer spacing, as observed in XRD results (Fig. 1f and Fig. S4). This suggests that Cl terminations removal and electron injection effectively weaken the interlayer interactions. The looser structure benefits subsequent single-layer exfoliation, making direct MXene nanosheet preparation feasible. SEM-EDS point and mapping analyses confirm that Cl elemental is almost entirely removed in the reduced samples, providing direct evidence for the complete elimination of -Cl terminations through surface reduction (Fig. S5-7).

Moreover, the lamellar structure of the $Ti_3C_2Cl_x$ as-synthesized reduced $Ti_3C_2$ MXenes is clearly revealed in the scanning transmission electron microscopy (STEM) images, as shown in Fig. 1g and h. STEM-EDS elemental mapping confirms the nearly complete removal of Cl element in the reduced $Ti_3C_2Cl$ sample, indicating highly efficient elimination of -Cl terminations. Atomic-resolution STEM imaging further reveals structural details at the atomic level. As shown in Fig. 1i, five distinct atomic layers are resolved for each $Ti_3C_2Cl_x$ layer, with the brighter atoms corresponding to Ti and the darker ones assigned to Cl. In contrast, the STEM image of the reduced $Ti_3C_2$ sample shows a significantly enlarged interlayer spacing and the absence of Cl atoms, providing direct and compelling evidence that Na metal effectively reduces and removes the Cl terminations (Fig. 1j). Notably, this enlarged interlayer spacing can facilitate the intercalation of various species, which may make it suitable in related applications.

The surface chemistry of the MXene samples was further investigated by X-ray photoelectron spectroscopy (XPS) (Figs. S8-12 and Tables S2-6). Figure 1k compares the Ti $2p$ spectra of MXenes synthesized under different conditions. Annealing leads to a stronger $TiO_2$ signal compared to pristine $Ti_3C_2Cl_x$, suggesting partial surface oxidation during thermal treatment, likely due to residual oxygen and moisture in the molten salt, even under an inert atmosphere. The Ti-C bonding, originating from the [$TiC_6$] octahedral building blocks in the $Ti_3C_2$ structure, appears as three distinct peaks in the Ti $2p$ spectrum. Notably, these peaks remain at nearly the same binding energy after annealing, suggesting that the thermal treatment itself does not alter the oxidation state of Ti. In contrast, for the reduced $Ti_3C_2$ MXene samples, a clear shift of these Ti-C characteristic peaks toward lower binding energies is observed with increasing reduction degree (Fig. 1k, l, and Tables S2-6). This downshift indicates electron transfer into the Ti 3*d* orbitals, effectively reducing the oxidation state of Ti. The extent of the shift correlates well with the reduction strength, providing direct spectroscopic evidence that Na-mediated reduction successfully injects electrons into the MXene lattice and modulates the Ti valence state. However, it should be noted that this peak shift is only suitable for qualitative analysis, as quantitative analysis is hindered by the varying degrees of surface oxidation among the different samples

**Preparation of Reduced $Ti_3C_2$ Nanosheets**

Given the significant increase in interlayer spacing observed in previous results, it is evident that the reduction process effectively mimics the effects of pre-intercalation steps typically required for MXene delamination.[31,32] This enhanced interlayer distance facilitates easier exfoliation of reduced MXene into nanosheets without the need for additional intercalation treatments. Consequently, reduced $Ti_3C_2$ (1:8, 550°C) with the largest interlayer spacing can be directly exfoliated into nanosheets via simple ultrasonication, streamlining the preparation process. However, other reduced $Ti_3C_2$ failed to be exfoliated despite their expanded interlayer spacing and subtle difference with $Ti_3C_2$ (1:8, 500°C) (Fig. S13). In contrast, $Ti_3C_2Cl_x$ can only be exfoliated after intercalation treatment using tetrabutylammonium hydroxide (TBAOH) (Figs. S14-16 and Table S7). These findings emphasize the significance of interlayer expansion in the exfoliation process and show the advantage of our reduction strategy.

The optical photograph in Fig. 2a demonstrates the successful exfoliation of reduced $Ti_3C_2$ into nanosheets, as evidenced by the pronounced Tyndall effect observed under laser illumination. This indicates a stable colloidal suspension, which is crucial for subsequent applications. TEM image (Fig. 2b) reveals that the reduced $Ti_3C_2$ nanosheets exhibit a highly uniform morphology with consistent particle sizes. The homogeneous distribution of these nanosheets suggests efficient delamination, leading to high-quality MXene nanosheets suitable for various applications. The selected-area electron diffraction (SAED) pattern shown in the inset of Fig. 2b confirms the hexagonal crystal structure of $Ti_3C_2$, indicating that the reduction process did not alter the fundamental crystal structure. This is essential for maintaining the intrinsic electronic and optical properties of the MXene. High-resolution TEM image (Fig. 2c) provides further insight into the atomic structure of the reduced $Ti_3C_2$ nanosheets. The lattice fringes are clearly indexed to the (100) plane, confirming the structural integrity of the reduced MXene nanosheets. Figure 2d presents the particle size distribution of the exfoliated nanosheets, highlighting their uniformity and narrow size range. This uniformity is critical for ensuring consistent performance in practical applications.

XRD analysis (Fig. 2e) compares the diffraction patterns of reduced $Ti_3C_2$ particles and exfoliated nanosheets. Both samples exhibit characteristic peaks corresponding to the hexagonal phase of $Ti_3C_2$ MXenes. However, the exfoliated nanosheets show only (00*l*) plane reflections, indicating a preferred orientation of the nanosheets. XPS (Fig. 2f and Fig. S17) was employed to compare the chemical states of $Ti_3C_2$ in both its nanoparticle and nanosheet forms. The XPS spectra of reduced $Ti_3C_2$ produced freshly and after 24h confirm that the Ti oxidation state remains unchanged between the two forms, indicating that the exfoliation process does not alter the chemical state of Ti (Figs. S17, 18, Tables S8, 9). Specifically, the presence of Ti-C bonds and the absence of Cl signals validate the complete removal of -Cl terminations in both samples (Fig. 2f and Fig. S17). The consistency in binding energies across both forms underscores the structural and chemical stability of the reduced $Ti_3C_2$ during sonication. Notably, the sonication process of reduced $Ti_3C_2$ MXene must be operated in ascorbic acid solution to avoid the oxidation caused by the high-energy damage and dissolved oxygen (Fig. S19). Also, a filtration and re-dispersion process to remove ascorbic acid will not cause obvious agglomeration (Fig. S20). The Raman results (Fig. 2g) show that although the reduction process suppresses the coordination ability of Ti atoms by electron injection, it does not completely eliminate their reactivity. Consequently, reduced MXene still forms a small amount of -OH and -O terminations in water, though at much lower levels than the pristine sample. This can be demonstrated by the Zeta potential results, where the zeta potential of reduced MXene nanosheets is approximately -7.1 mV (Fig. 2h). Compared to $Ti_3C_2Cl_x$ (-37.7 mV) or HF-etched $Ti_3C_2T_x$ nanosheets (-38.2 mV), which typically exhibit much higher negative charges, the lower zeta potential here indicates a significant reduction in surface termination density, corroborating the effectiveness to the suppression of surface terminations.[33]

**Electronic Behavior Analysis**

The electronic properties of pristine $Ti_3C_2Cl_x$ and reduced $Ti_3C_2$ samples prepared under different conditions were systematically investigated through Hall effect measurements (Fig. 3a, Fig. S21, and Table S11). These measurements assessed key parameters, carrier density, mobility, conductivity, and Hall coefficient, revealing how surface termination removal and electron injection influence MXene's electronic behavior (Fig. 3a and Table S11). The results clearly demonstrate that surface termination removal ($Ti_3C_2$, (1:2, 500°C)) leads to only a modest increase in carrier concentration, approximately 16.82%, which is primarily due to the elimination of the electron-withdrawing effect caused by -Cl terminations. However, this treatment has a more pronounced effect on carrier mobility, which increases by 35.53%, which may be attributed to the reduction in electron scattering caused by -Cl terminations.

When the reduction process is further intensified, specifically at higher Na ratios (1:8) and elevated temperatures (500°C and 550°C), the enhancement in electrical performance becomes significantly more pronounced. Compared to $Ti_3C_2Cl_x$, for reduced $Ti_3C_2$ (1:8, 500°C), the carrier density, the mobility, and the overall conductivity increase by 121.17%, 86.64%, and 313.23%, respectively. The reduced $Ti_3C_2$ (1:8, 550°C) exhibits an extraordinary increase in carrier concentration by 392.49%, in mobility by 162.85%, and in conductivity by a significant 1196%. These results strongly indicate that as the degree of reduction increases, the amount of electron injection into the MXene lattice becomes substantial, leading to a dramatic rise in the number of free charge carriers and consequently enhancing the material's intrinsic conductivity.

To further understand the underlying mechanism, ultraviolet photoelectron spectroscopy (UPS) was conducted on both the pristine $Ti_3C_2Cl_x$ and the most effectively reduced $Ti_3C_2$ (1:8, 550°C). As shown in Fig. 3b, c, the work function of the reduced MXene decreases from 4.25 eV to 3.79 eV, which is primarily attributed to the large amount of electron injection during the reduction process. The comparison of Fermi edge features between the two samples reveals a much steeper slope in the reduced $Ti_3C_2$, indicating a significantly higher concentration of free electrons near the Fermi level (Fig. 3d, e). This sharp increase in available charge carriers results in a stronger photoelectron signal upon ultraviolet excitation, directly supporting the conclusion that the reduction process greatly enhances the density of states near the Fermi level.

To gain deeper insight into how reduction affects the electronic structure of MXene, theoretical calculations were performed (Figs. S22-25 and Table S12). As illustrated in Fig. 3f and g, the -Cl terminations in pristine $Ti_3C_2Cl_x$, due to their high electronegativity, effectively draw electrons away from the outer Ti atoms, resulting in lower electron density in these regions. Inner Ti atoms, however, remain largely unaffected by the presence of -Cl terminations. Upon complete removal of the terminations, the suppression of electron density at the outer layers is lifted, leading to a noticeable increase in electron availability (Fig. 3h and i). When additional electrons are injected into the system, both outer and inner Ti atoms experience a significant enhancement in electron density, demonstrating that the reduction not only impacts the surface but also penetrates the inner structure (Fig. 3j and k).

Further analysis of the energy band structure and DOS reveals that in pristine $Ti_3C_2Cl_x$, the presence of -Cl terminations causes the electrons near the Fermi level to be more dispersed (Fig. 3l). After termination removal, these electrons begin to accumulate closer to the Fermi level, increasing the local charge density and improving conductivity (Fig. 3m). With continued electron injection, the electron density at the Fermi level increases even further, leading to a dramatic enhancement in electrical properties (Fig. 3n). These theoretical observations align closely with the experimental data, providing a unified explanation for the observed improvements in carrier concentration, mobility, and overall conductivity in reduced MXene materials.

**Photothermal Property Study**

Based on our previous findings, the surface reduction process significantly enhances the free carrier concentration in MXenes. This increase in electron density is expected to have a profound impact on their optical and electronic properties, particularly those related to photothermal conversion. Thus, we investigated how the surface reduction strategy affects their photothermal behavior of MXenes.

The photothermal conversion performance of pristine $Ti_3C_2Cl_x$ and reduced $Ti_3C_2$ (550°C, 1:8) nanosheets was evaluated using ultraviolet-visible (UV-Vis) spectroscopy and laser irradiation experiments. As shown in Fig. 4a and b, both materials exhibit absorption in the near-infrared region, which is essential for efficient photothermal conversion. However, a notable blue shift in the absorption peak is observed for the reduced $Ti_3C_2$ compared to the pristine $Ti_3C_2Cl_x$ (Fig. S26). This blue shift suggests a change in the LSPR wavelength, which may be attributed to the increased free electron concentration and the elimination of -Cl terminations. Figure 4c compares the mass extinction coefficient at 808 nm for both samples. The reduced $Ti_3C_2$ exhibits a much higher extinction coefficient (18.04 L g$^{-1}$ cm$^{-1}$) than the pristine $Ti_3C_2Cl_x$ (7.56 L g$^{-1}$ cm$^{-1}$). This confirms the enhanced light absorption capability after reduction, which originates from the increased electron concentration.

Photothermal heating curves under 808 nm laser irradiation (Fig. 4d and e) show that the reduced $Ti_3C_2$ achieves a significantly higher temperature rise at the same power density and concentration. For instance, at 50 μg ml$^{-1}$, the reduced sample reaches ~65°C, while the pristine sample only reaches ~45°C (Video S1). Moreover, compared to gold nanoparticles (AuNPs), reduced $Ti_3C_2$ shows a greater advantage in temperature rise (Video S2). This superior photothermal performance can be attributed to both the enhanced light absorption due to increased free electron concentration and the improved charge transport properties resulting from the removal of surface scattering centers. Photothermal conversion efficiency was calculated based on the temperature rise and cooling rate (Fig. 4f and g, Fig. S27). The reduced $Ti_3C_2$ shows a high efficiency of 92.63%, indicating an enhancement of approximately 51.3% compared to 61.22% for the pristine $Ti_3C_2Cl_x$. Moreover, stability tests over five laser on/off cycles (Fig. 4h and i) demonstrate that both the pristine $Ti_3C_2Cl_x$ and reduced $Ti_3C_2$ samples exhibit good photothermal stability, maintaining consistent heating and cooling profiles without significant degradation. Notably, this obtained PTCE of reduced $Ti_3C_2$ is the highest reported value for MXene materials at 808 nm irradiation (Fig. 4j, Table S13).

**Application exploration on the photothermal antibacterial woundplast**

To explore the biomedical potential of reduced $Ti_3C_2$ MXene, its cytotoxicity and biocompatibility were first evaluated using human umbilical vein endothelial cells (HUVEC) and Raw 264.7 cells through standard cell toxicity assays and live/dead cell viability staining (Figs. S28-30). The results showed negligible cytotoxicity across all tested concentrations, with over 90% cell viability maintained even at the highest MXene loading (200 μg/mL), demonstrating excellent biocompatibility and laying the foundation for further biomedical applications.

Based on these findings, a photothermal antibacterial woundplast was fabricated using reduced $Ti_3C_2$ MXene (1:8, 550°C) (Fig. S31, 32). Considering the safe temperature range for photothermal therapy (42-60°C), samples were irradiated with an 808 nm laser (1.0 W/cm$^2$), and real-time temperatures were recorded using infrared imaging (Fig. 5a-c). The 20 μg/mL group reached ~60°C within 1 min, achieving effective photothermal performance while minimizing thermal damage risk. Thus, 20 μg/mL was selected as the optimal loading concentration. Four control groups, AuNPs, phosphate buffered saline (PBS), pristine $Ti_3C_2Cl_x$, and reduced $Ti_3C_2$, were prepared and tested under identical conditions (Figs. S33-36). Only the reduced $Ti_3C_2$ group exhibited rapid heating; no significant temperature rise was observed in AuNPs or $Ti_3C_2Cl_x$ groups (see Videos S3, 4).

For antimicrobial evaluation, *Staphylococcus aureus* was selected as the model bacterial strain. As shown in Fig. 5d, the colony-forming unit (CFU) assay reveals that woundplast loaded with reduced $Ti_3C_2$ exhibits excellent antibacterial performance under 808 nm laser irradiation, with nearly no colonies observed on the agar plate. In contrast, other groups (PBS, AuNPs, and $Ti_3C_2Cl_x$) show minimal inhibition of bacterial growth, confirming the critical role of the photothermal effect in enhancing antibacterial efficacy. Figure 5e presents the quantitative analysis of antibacterial rates based on CFU counting. Under NIR irradiation, the reduced $Ti_3C_2$ group exhibited an antibacterial rate of 91.39%, significantly higher than that of the $Ti_3C_2Cl_x$ or AuNPs groups (Fig. S37). In contrast, no notable antibacterial effect was observed in the absence of laser irradiation, confirming that the photothermal effect is the primary mechanism responsible for bacterial inactivation. To assess the long-term antibacterial performance, bacterial growth was monitored over 12 hours *via* $OD_{600}$ measurements (Fig. 5f). The reduced $Ti_3C_2$ group under laser irradiation showed consistently low optical density values, indicating effective suppression of bacterial proliferation. In comparison, all other groups exhibited rapid bacterial regrowth within a few hours, highlighting the superior and lasting antimicrobial capability of the MXene-based photothermal treatment.

Notably, while the *in vitro* results demonstrate strong antibacterial efficacy, it is important to consider the influence of baseline body temperature (~37°C) *in vivo*, which may affect the local thermal response during photothermal therapy. Therefore, precise regulation of MXene loading will be critical in future *in vivo* studies to ensure both therapeutic effectiveness and thermal safety when applied to living tissues.

**Conclusion**

In summary, a novel strategy for sodium-mediated surface reduction in molten salts is developed to effectively remove electronegative Cl terminations and simultaneously inject electrons into the 3*d*-orbitals of Ti, thereby significantly improving the photothermal performance of $Ti_3C_2$ MXene and achieving a photothermal conversion efficiency of 92.36% under 808 nm laser irradiation. This represents the highest efficiency reported for MXenes at this widely used wavelength so far. The reduced MXene not only exhibits excellent photothermal performance but also has highly efficient antibacterial activity at ultralow loading levels, achieving a bacterial kill rate of more than 91.39% in a model woundplast application. This demonstrates the key role of surface chemistry and electron concentration in regulating the photothermal properties of MXene and the great potential of reduced MXene in applications requiring high electron concentration, such as energy storage, sensors, and electromagnetic interference shielding.

**Author Contributions**

Y. Z. supervised and initiated the work; H. M. D. conducted the main experiment and analyzed results with X. T.; Y. Z., H. M. D., and X. T. wrote the manuscript; H. M. D. and X. T. contributed equally to this work. All coauthors read and commented on successive drafts of the manuscript.

**Acknowledgment**

This study was funded in part by the InnoHK initiative of the Innovation and Technology Commission of the Hong Kong Special Administrative Region Government, in part by the Wearable Intelligent Sensing Engineering (WISE) of Hong Kong Centre for Cerebro-cardiovascular Health Engineering (COCHE), and in part by Novel Fluorescent Microscopy Platforms for 3D Imaging of Hong Kong Centre for Cerebro-cardiovascular Health Engineering (COCHE), the City University of Hong Kong (project number 9380160) and its Institute of Digital Medicine (project number 9229502, 9229501-18-ZY), Hong Kong government's Global STEM Professorship, and Shenzhen Medical Academy of Research and Translation (SMART, project number B2402007).

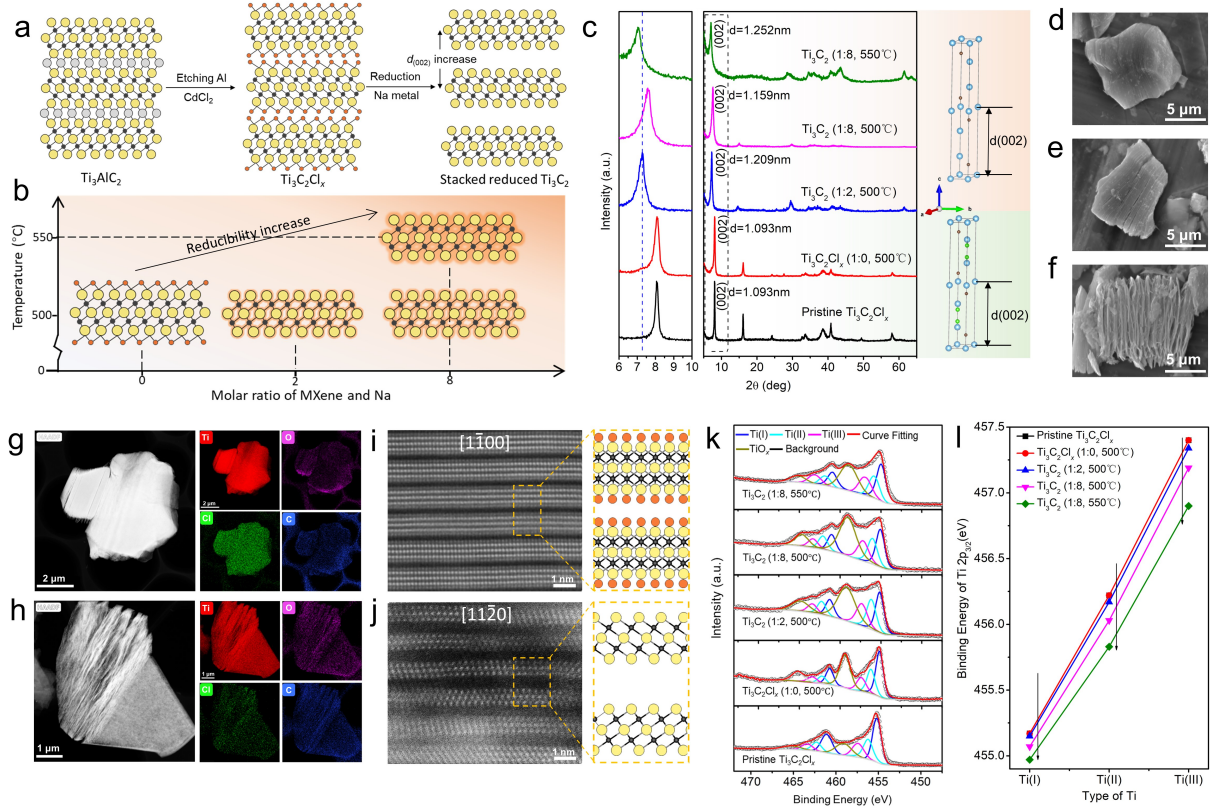

**Fig. 1.** Synthesis of reduced MXene through a Na-mediated molten salt reduction strategy. (a) Schematic illustration of the synthesis process of reduced $Ti_3C_2$ MXenes; (b) The regulation of the conditions of reduction reactions; (c) X-ray diffraction (XRD) patterns of pristine $Ti_3C_2Cl_x$ MXene and reduced $Ti_3C_2$ MXene synthesized under different reaction conditions; Scanning electron microscopy (SEM) image of pristine $Ti_3C_2Cl_x$ (d), annealed $Ti_3C_2Cl_x$ in molten salts (e), and reduced $Ti_3C_2$ (1:8, 550°C) (f); High-angle annular dark field (HAADF) transmission electron microscopy (TEM) image of pristine $Ti_3C_2Cl_x$ (g) and reduced $Ti_3C_2$ (1:8, 550°C) (h), and their corresponding energy dispersive spectroscopy (EDS) mapping of Ti, Cl, O, and C elements ; (i) Scanning transmission electron microscopy (STEM) of pristine $Ti_3C_2Cl_x$, showing its atomic structure with the beam along the $[1\bar{1}00]$ axis. (j) STEM image of reduced $Ti_3C_2$ (1:8, 550°C), showing the atomic structure with the beam along the $[11\bar{2}0]$ after removal of Cl terminations. (k) X-ray photoelectron spectroscopy (XPS) results of pristine $Ti_3C_2Cl_x$, annealed $Ti_3C_2Cl_x$ and reduced $Ti_3C_2$ MXenes synthesized under different conditions. (i) Changes in the binding energy of varying Ti within different MXenes.

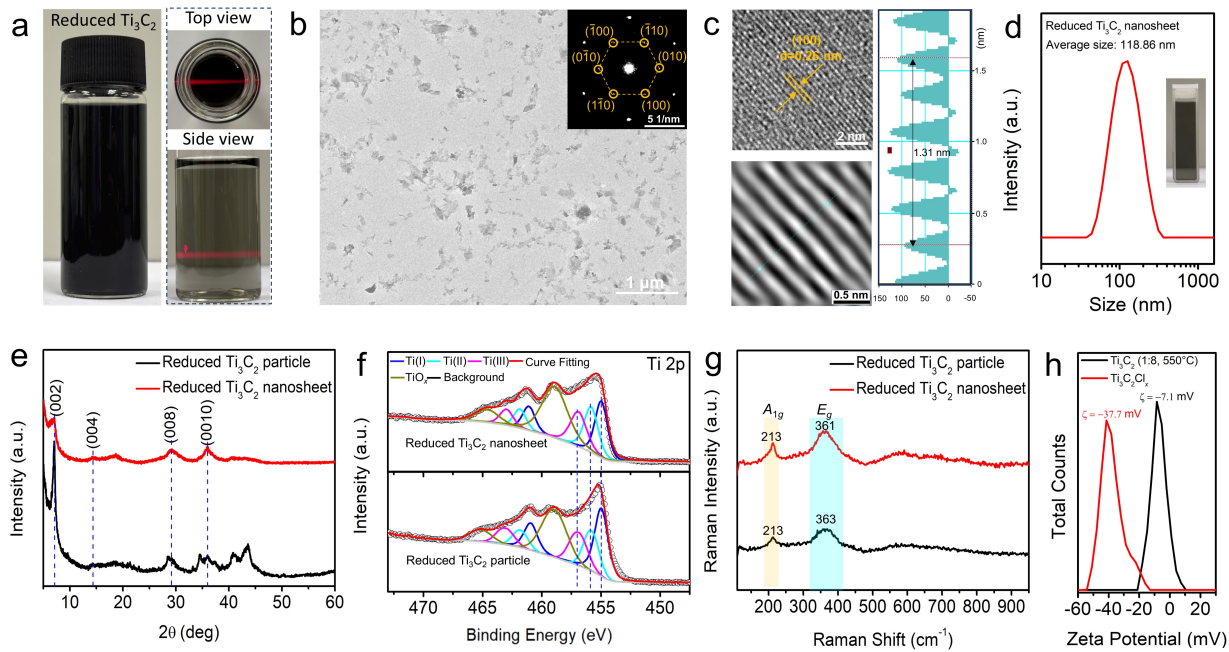

**Fig. 2.** Preparation of reduced $Ti_3C_2$ (1:8, 550°C) nanosheets. (a) Optical photograph of the reduced $Ti_3C_2$ nanosheet suspension with a high concentration (left) and a diluted concentration (right), showing the Tyndall effect under laser illumination. (b) TEM image of the reduced $Ti_3C_2$ nanosheets, revealing uniform morphology and particle size distribution, and the inset shows the selected-area electron diffraction (SAED) pattern corresponding to the hexagonal structure of $Ti_3C_2$, indicating preserved crystallinity. (c) High-resolution TEM lattice image indexed to the (100) plane of reduced $Ti_3C_2$ and its corresponding Fourier transform of the selected area (left), and the measured lattice spacing based on the Fourier transform. (d) Size distribution of the exfoliated nanosheets. (e) XRD patterns of both reduced $Ti_3C_2$ MXene particles and their corresponding exfoliated nanosheets. (f) XPS spectra of reduced $Ti_3C_2$ particles and nanosheets. (g) Raman spectra of reduced $Ti_3C_2$ particles and nanosheets. (h) Zeta potential measurement in neutral deionized water of $Ti_3C_2Cl_x$ nanosheets and reduced $Ti_3C_2$ nanosheets.

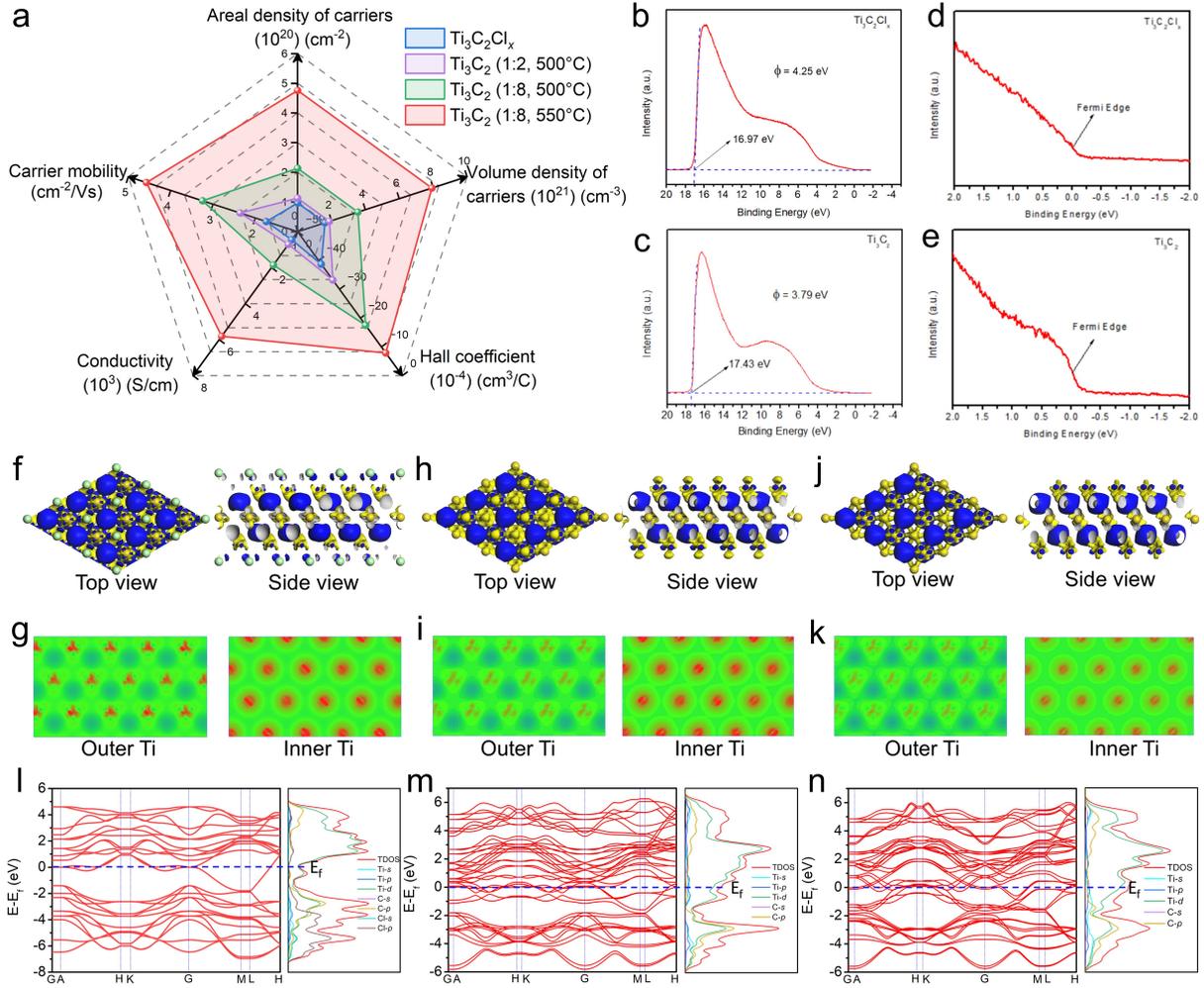

**Fig. 3.** Electronic structure analysis. (a) Rada map of the electronic structure information of different MXenes based on the test of the Hall effect measurement, showing the carrier density, mobility, conductivity, and Hall coefficient; Ultraviolet photoelectronic spectra of $Ti_3C_2Cl_x$ (b) and reduced $Ti_3C_2$ (c) and their corresponding Fermi edges (d,e), respectively. Three-dimensional differential charge density images of $Ti_3C_2Cl_x$ (f), reduced $Ti_3C_2$ without terminations (h), and reduced $Ti_3C_2$ injected with electrons (j) from the top and side views. In-plane 2D charge density images of $Ti_3C_2Cl_x$ (g), reduced $Ti_3C_2$ without terminations (i), and reduced $Ti_3C_2$ injected with electrons (k), showing the charge distribution of outer and inner Ti, where red represents the positive charge, green represents the neutral background. Energy band structures and DOS of $Ti_3C_2Cl_x$ (i), reduced $Ti_3C_2$ without terminations (m), and reduced $Ti_3C_2$ injected with electrons (n).

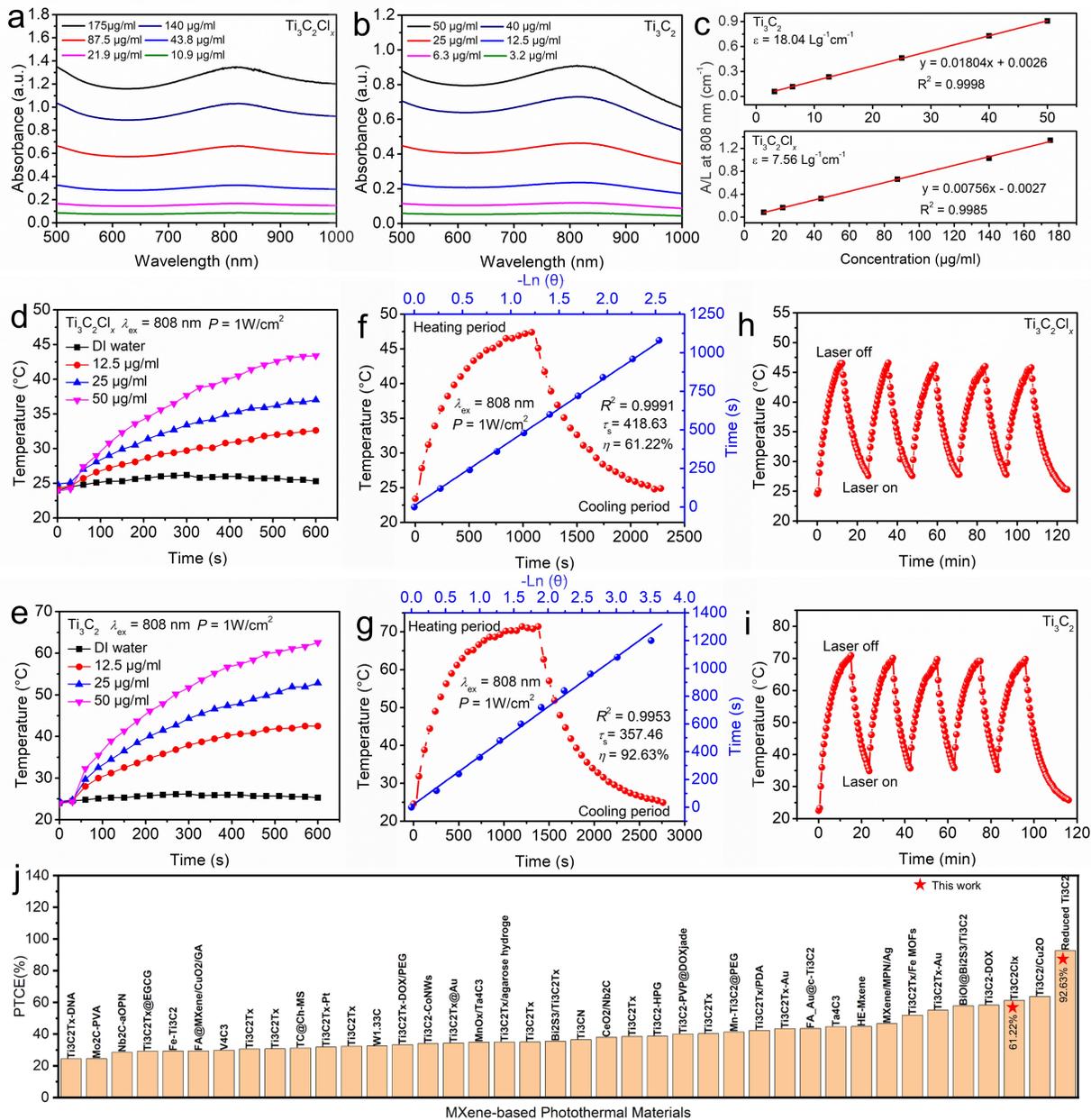

**Fig. 4.** Photothermal-conversion performance of pristine Ti$_3$C$_2$Cl$_x$ and reduced Ti$_3$C$_2$ (550°C, 1:8) nanosheets. (a) Ultraviolet-visible Spectroscopy (UV-Vis) absorbance spectra of aqueous suspensions of dispersed Ti$_3$C$_2$Cl$_x$ nanosheets at varied concentrations (10.9, 21.9, 43.8, 87.5, 140, and 175 μg mL$^{-1}$) (b) UV-Vis absorbance spectra of aqueous suspensions of dispersed reduced Ti$_3$C$_2$ nanosheets at varied concentrations (3.2, 6.3, 12.5, 25, 40, and 50 μg mL$^{-1}$); (c) Mass extinction coefficient of Ti$_3$C$_2$Cl$_x$ and reduced Ti$_3$C$_2$ nanosheets at 808 nm. Normalized absorbance intensity at λ = 808 nm divided by the characteristic length of the cell (*A/L*) at varied concentrations. Photothermal heating curves of aqueous suspensions of dispersed Ti$_3$C$_2$Cl$_x$ (d) and reduced Ti$_3$C$_2$ (e) nanosheets under irradiation of an 808 nm laser at varied concentrations (12.5, 25, and 50 μg mL$^{-1}$). Calculation of photothermal conversion efficiency at 808 nm for Ti$_3$C$_2$Cl$_x$ (f) and reduced Ti$_3$C$_2$ (g) nanosheets. Red line: temperature rise during laser irradiation and temperature down after turning off the laser; blue line: cooling phase with fitted time constant ($\tau_s$) for heat dissipation. Stability test over five on/off laser cycles (1 W cm$^{-2}$) for Ti$_3$C$_2$Cl$_x$ (h) and reduced Ti$_3$C$_2$ (i) nanosheets. (j) Summary of the PTEC of different MXene-based nanosystems.

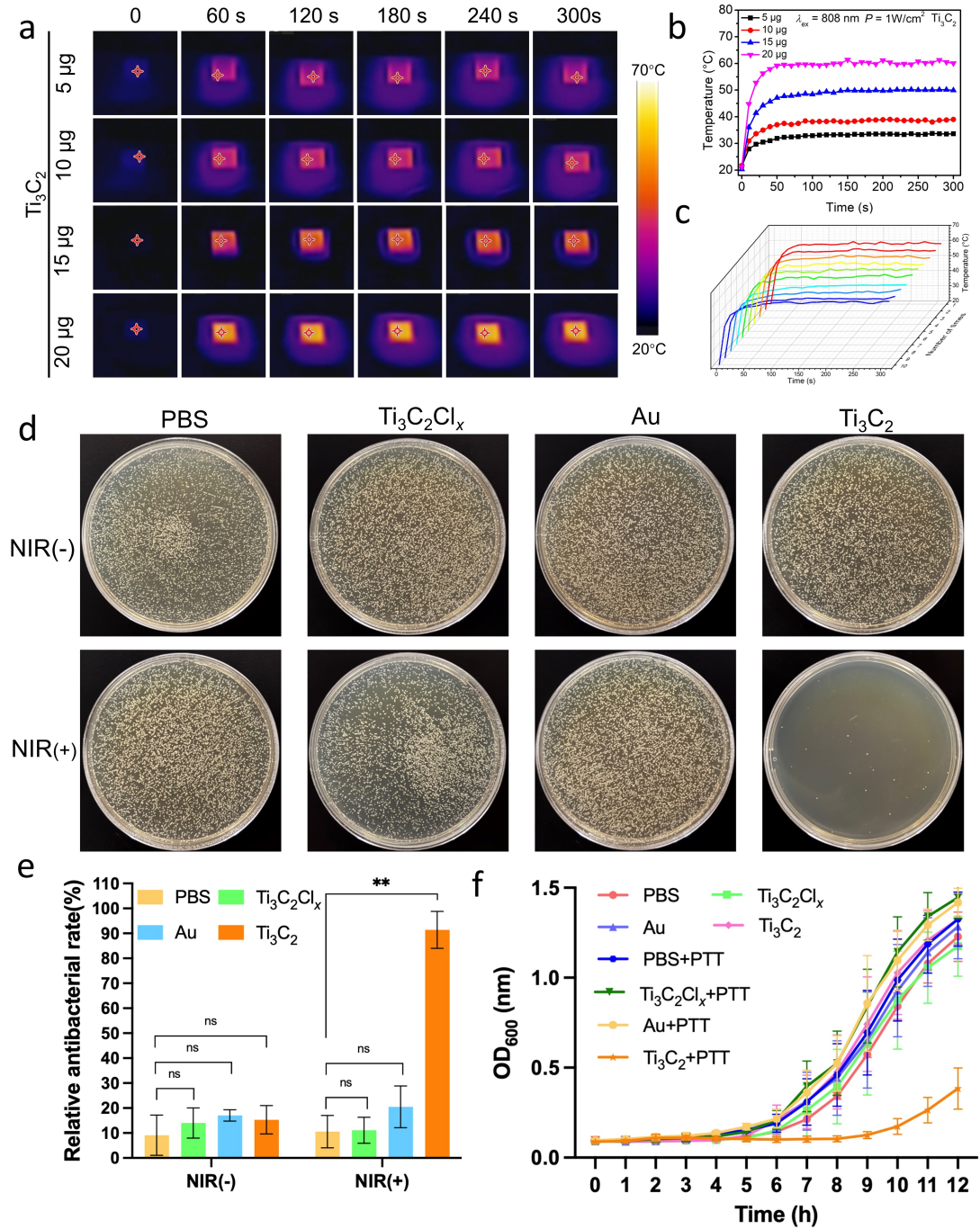

**Fig. 5.** The photothermal antibacterial effect of woundplast loaded with reduced $Ti_3C_2$ nanosheets (1:8, 550°C). (a) Infrared thermal images of woundplast loaded with reduced $Ti_3C_2$ at different concentrations under 808 nm laser irradiation with a power density of 1 W/cm$^{-2}$. (b) Photothermal heating curves of woundplast loaded with reduced $Ti_3C_2$ at different concentrations under 808 nm laser irradiation with a power density of 1 W/cm$^{-2}$. (c) Photothermal cycling stability of woundplast loaded with reduced $Ti_3C_2$ at different concentrations under 808 nm laser irradiation with a power density of 1 W/cm$^{-2}$. (d) Colony-forming unit (CFU) assay results of antibacterial treatments using woundplast loaded with different photothermal agents. (f) Relative antibacterial rates of different treatment groups. (g) Bacterial growth curves over 12 hours.